\def\preprint{1}		
\def\comment#1{}
\if\preprint1
	\documentclass[usenatbib]{mn2e}
        \usepackage{natbib}
	\usepackage{times}
	\usepackage{graphicx}
	\usepackage{calc}
\else
	\documentstyle[astrop-bib,referee,times]{mn}
	\newcommand{\includegraphics}[1]{}
\fi

\def\oversim#1#2{\lower0.5pt\vbox{\baselineskip0pt \lineskip-0.5pt
     \ialign{$\mathsurround0pt #1\hfil##\hfil$\crcr#2\crcr\sim\crcr}}}
\def\lsim{\mathrel{\mathpalette\oversim<}}    

\title[AGB Supernovae]
{Low-mass  supernovae in the early Galactic halo: source of the double  
$r$/$s$-process enriched halo stars?}
\author[A.A. Zijlstra]
       {Albert~A.~Zijlstra \thanks{E-mail: \tt a.zijlstra@umist.ac.uk}\\
	$^1$UMIST, Department of Physics, P.O. Box 88, Manchester M60 1QD, UK\\
}
\pubyear{2003}

\begin{document}

\maketitle

\begin{abstract}

Several stars at the low-metallicity extreme of the Galactic halo
([Fe/H]$\,=-2.5$) show strong enhancements of both s-process and
r-process elements.  The presence of s-process elements in
main-sequence stars is explained via mass transfer from an AGB
companion star in a binary system.  r-Process elements originate in
type-II supernovae and also require mass transfer. It is however
unclear how pollution by both an AGB star and a supernova could have
occured. Here I show that the initial--final-mass relation steepens at
low metallicity, due to low mass-loss efficiency. This may cause the
degenerate cores of low-$Z$, high-mass AGB stars to reach the
Chandresekhar mass, leading to an Iben \& Renzini-type-1.5 supernova.
Such supernovae can explain both the enhancement patterns and the
metallicity dependence of the double-enhanced halo stars.  Reduced
mass loss efficiency predicts more massive remnants in metal-poor
globular clusters.  The evidence for a high $M/L$ population in the
cores of globular clusters is briefly discussed.

\end{abstract}

\begin{keywords}
stars:  stars: AGB and post-AGB
 -- stars: mass-loss
\end{keywords}

\section{Chemical peculiarities in halo stars}

The metallicity distribution of the stars in the Galactic halo ranges
from [Fe/H]$\,=-1$ for the most metal-rich stars, to [Fe/H]$=-5.3$ at
the minimum (HE 0107$-$5240: \citet{CBB02}). The first
halo stars formed from almost unprocessed primordial
gas. A short formation time is favoured, because chemical enrichment
due to supernovae is rapid. The Milky Way disk formed out of
gas pre-enriched to about [Fe/H]$\approx-1 $. The pre-enrichment may
have been due to the Bulge formation \citep{Renzini2003} but the most
metal-rich stars in the halo show that this level of enrichment was
already reached by the end of the halo formation. The global metallicity
of the Universe reached [Fe/H]$\,\approx-1 $  at $z=3$ \citep{Renzini2003}.

Some halo main-sequence stars show evidence for abundance alterations.
Because main-sequence stars should not have significantly altered
their surface abundances, and pollution from planetary companions is
unlikely given the dearth of planets at subsolar metallicity
\citep{BSIM03}, pollution from a companion star by mass transfer
is considered most likely. In some cases high overabundances of
s-proces elements (e.g. Pb) are found \citep{ANR00,EGJP01,LGC03}.
These elements are formed in Asymptotic Giant branch (AGB) stars
\citep{EGJP03}, which first appeared $10^8$\,yr  after the halo
formation.  Other metal-poor stars show evidence for large enhancements of
r-process elements, a.o. europium \citep{SCI00} and gold
\citep{CSB02}. These form in type-II supernovae, and are also
explained via pollution from a companion star.

The discovery that several stars show enhancements of {\it both}
r-process and s-process elements \citep{Hill2000,CCQW03} is puzzling,
as they require pollution from both an AGB star and a supernova.

In this paper we will present a calculation which indicates that at
low metallicity, inefficient mass loss may allow the degenerate cores
of AGB stars to reach the Chandresekhar mass. This would lead to a
AGB supernovae, polluting low-mass companions by both s-process
elements (enriched in the AGB envelope) and r-process elements formed
in the subsequent explosion.  The next section discusses the
double-polluted stars.  Section 3 discusses AGB mass loss at low $Z$,
and Section 4 derives initial-final masses for these stars and
discusses the possibility of AGB supernovae.

\section{Double polluted halo stars}

Three stars are known to show the double enhancements: HE2148$-$1247
\citep{CCQW03}, CS22948$-$027 \citep{Hill2000,SP01} and CS29497$-$034
\citep{Hill2000}, with [Fe/H]=$-2.3$,$-2.45$, and $-2.90$
respectively. CS22898$-$027 \citep{SP01,Aoki02}, at [Fe/H]$=-2.25$,
may also show double enrichment.  The metallicities are at the lower
range of the gaussian halo distribution \citep{RS91}.  The process
leading to the double enhancement appears to require [Fe/H]$\,<-2$.  A
lower limit to the metallicities of the double enriched stars is not
evident.  Although there is a tail in the halo distribution extending
to [Fe/H]$=-4$ \citep{RS91} or lower \citep{CBB02}, there are few
stars in this tail.

\citet{CCQW03} discuss possible scenarios for double enhancements. 
A triple system including one massive pre-supernova and one pre-AGB
star, plus the remaining low-mass star, appears very unlikely.  They
therefore suggest reverse mass-transfer from the low-mass star to the
post-AGB white dwarf, leading to accretion-induced collapse, and 
a type-Ia supernova.  A third star in
a close orbit around the AGB primary may instead be considered for the mass
donor, leading to a more traditional type-Ia scenario, but this again
leads to an unlikely close triple system. There is also no evidence
that type-Ia SNe manufacture significant amounts of r-process
elements, arguing against both these scenarios.

Pollution from an AGB companion (s-process elements) occurs via wind
accretion. The peak transfer efficiency ocurs for binaries with
periods around 3000 days \citep{PKLT03}.  \citet{HEPT95} stress the
additional importance of a common envelope phase in mass transfer, but
orbital eccentricities show that all but the closest binaries ($d <
1\,\rm AU$) avoid a common envelope \citep{KTL2000,PKLT03}.  Roche
lobe overflow is also unlikely \citep{Han2002}, firstly because it
tends to happen on the first giant branch (before the onset of the
s-process enhancements) and secondly because it is unstable for donor
stars more massive than the secondary, as would have been the case
here. For the more likely wider orbits, the wind interaction has
little or no effect on the primary AGB star, which continues to evolve
as in the case of a single star. 

The wide orbit argues against the accretion-induced collapse, because
of the need to transfer a large amount of mass from a distant low-mass
main-sequence star in order to reach the Chandresekhar mass.
A scenario which does not require either a triple
system or reverse mass transfer, and explains the metallicity dependence
of the double-enrichment process, could be of interest.

\section{AGB mass loss at low metallicity}

\subsection{Mass loss expectations}

Stars on the AGB burn helium and hydrogen in shells around their inert
C/O core. Mass loss increases during the AGB: once the mass-loss rate
in the wind significantly exceeds the nuclear burning rate ($\sim
10^{-7}\,\rm M_\odot\,yr^{-1}$), the evolution comes to a sudden end
\citep{Willson2000}. The remaining envelope is quickly removed and
nuclear burning will cease. The C/O core at this point forms
the subsequent white dwarf.

The dependency of the mass loss on the stellar parameters is not well known.
Several formalisms have been proposed in the literature, depending typically
on the radius, mass and luminosity of the star. The one used most
extensively is the \citet{Bloecker95}  relation

\begin{equation}
\dot M = -4.8\, 10^{-9} \frac{LR}{M} \frac{L^{2.7}}{M} \,\rm M_\odot\,yr^{-1},
\end{equation}

\noindent with quantities in solar units.   Such relations are generally 
derived  for Galactic stars. An explicit metallicity dependence is not
included.

The mass loss of AGB stars is a two-step process. First, pulsations
extend the atmosphere and drive a small mass loss. Secondly, dust
forms in the extended atmosphere: radiation pressure on the dust now
drives the large mass-loss rates observed.  At low metallicity, the
efficiency of the dust formation is reduced and this will limit the
mass-loss rates.  \citet{BW91} derive mass-loss rates at low $Z$ from
theoretical pulsation models. They find that at low metallicity
([Fe/H]$<-1$) the dust does not play a significant role and the wind
becomes purely pulsation driven. Stars with lower metallicity are
found to have lower mass-loss rates.  This will allow the core to grow
to a larger mass and reach higher final masses.

The AGB wind depends on the stellar radius.  As low-metallicity stars
have smaller radii (\citet{Iben84} gives for AGB stars the relation $R\propto
Z^{0.088}$), this implies a further metallicity dependence of the mass
loss, even for dust-free, pulsation driven winds.

\subsection{Observational Evidence}

\begin{figure*}
\includegraphics[width=\textwidth,clip=true]{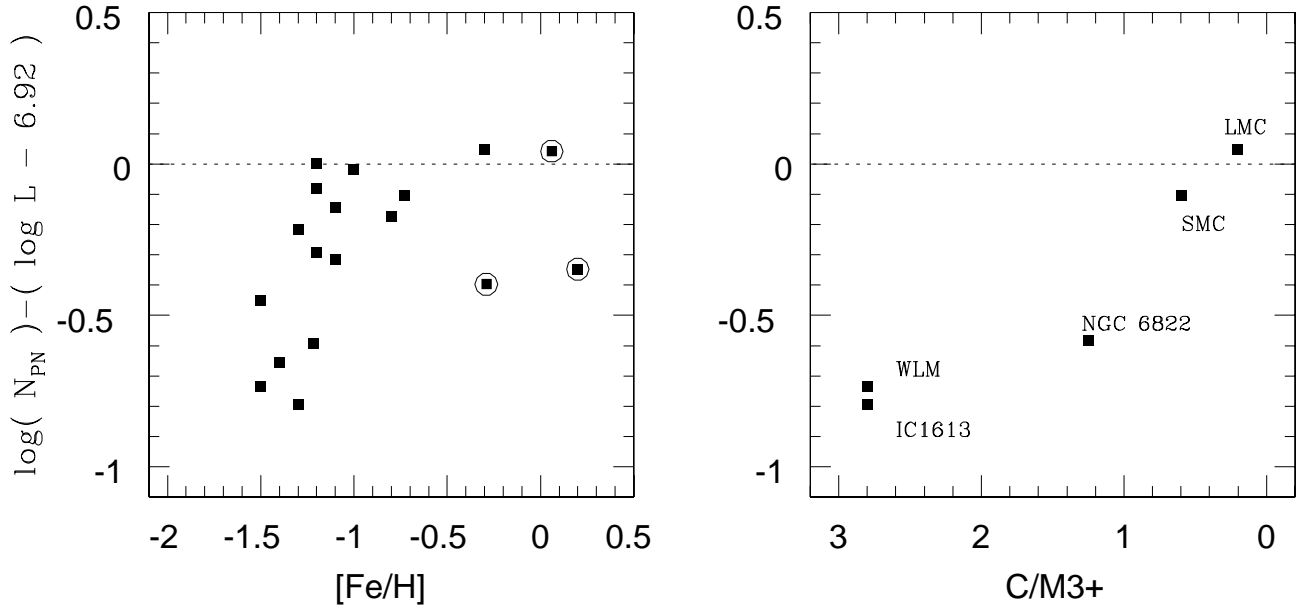}
\caption{\label{pne} Left: The ratio between number of planetary nebulae
and luminosity of the parent stellar population ($y$-axis), as
function of metallicity ($x$), for nearby galaxies
\citep{Magrini03}. Encircled points indicate spiral galaxies affected
by internal extinction. Right: The same ratio, as function of the ratio of
carbon over M-type stars on the AGB, a tracer of the metallicity of
the AGB population}
\end{figure*}
  
There are few direct observations of mass loss at low metallicity.
The best data set comes from ISO observations of AGB stars in the
Magellanic Clouds \citep{TLW99}, from which both mass-loss rates and
luminosities were derived \citep{LGK99}. Comparison with Galactic
stars is not straightforward, because within the Galaxy distances tend
to be poorly known and the luminosities are uncertain. The ISO data
shows that both LMC and SMC stars reach similar mass-loss rates to
those shown by Galactic stars near the tip of the AGB
\citep{Loon2000}, although the unknown dust-to-gas ratios and
expansion velocities could hide some difference.  The luminosity of
the LMC and SMC stars tends to be high, reaching up to $L=5 \times
10^4\,\rm L_\odot$.

Mass loss at even lower metallicities  ([Fe/H]$<-0.7$) is very poorly
studied. However, the superwind phase will lead to the
formation of a planetary nebula (PN) and these can be observed to very
large distances.  There is a good relation between the number of PNe
and the luminosity of the host galaxy: for the most populated systems
(Bulge, LMC), $\log [N({\rm PN)}] \approx \log L_V/L_\odot -
6.9$ \citep{Magrini02}.  However, local galaxies with low $Z$ show
some evidence for a deficiency in the number of PN
\citep{Magrini03}. This becomes apparent only for [Fe/H]$<-1$
(Fig. \ref{pne}), well below the LMC or SMC metallicity. Plotting the
deficiency as function of ratio of C-type over M-type AGB stars
(Fig. \ref{pne}, right panel) shows a clear relation: this ratio
is an effective tracer of the metallicity of the AGB population
\citep{Groen99}.

The deficiency of PNe suggests that the mass-loss rates do not reach
as high values for metallicities [Fe/H]$\,\lsim-1$. Lower peak
mass-loss rates will give rise to less dense PNe which will fade
faster. Conversely, the lack of a relation at [Fe/H]$\,>-1$ suggests
that all these AGB stars reach similar mass-loss rates (although not
necessarily at the same luminosities) regardless of metallicity.  This
change of behaviour at [Fe/H]$\,=-1$ is in excellent agreement with
the predictions of \citet{BW91}.

\section{AGB supernovae}

We can estimate the effect of metallicity on the initial--final-mass.
The goal is to obtain qualitative estimates: approximate, simple
relations will suffice. In a more detailed calculation, more
complicated relations could be substituted but these introduce their
own uncertainties.  The largest uncertainties comes from the mass-loss
estimates, and not including hot bottom burning.

The relation between the luminosity corresponding to a certain
mass-loss rate, and the metallicity, is taken from \citet{BW91}.
Between $Z=Z_\odot$ and $Z=0.1\times Z_\odot$, this luminosity
increases by a factor of 1.3, in part because the wind is no longer
dust driven. At lower Z, the luminosity required to support the same
mass-loss rate scales as \citep{BW91} :

\begin{equation}
\log L(Z)/L(Z_\odot) = 0.12 -0.13 \log Z/Z_\odot; 
\quad  Z/Z_\odot \le 0.1.
\end{equation}

The effect on the final mass of the star is derived from the
core mass--luminosity relation, which we take
from  \citet{BS88}:

 \begin{equation}
  \frac{L}{L_\odot} = 59250 \left(\frac{M_{\rm c}}{M_\odot}-0.495\right)
\end{equation}

The increase of the final mass is very small for low core mass,
but becomes significant for core masses corresponding to the heaviest
white dwarfs.  We use an approximate present initial--final mass
relation corresponding to solar metallicity:

\begin{equation}
M_{\rm f} = 0.5+\frac{1}{12}M_{\rm i}.
\end{equation}

\noindent The true relation is shallow at low core mass and 
steepens at high core mass \citep{VW93}, so this is a significant
simplification which may be more realistic at high than at low $M_{\rm
c}$.

\begin{figure*}
\includegraphics[width=\textwidth]{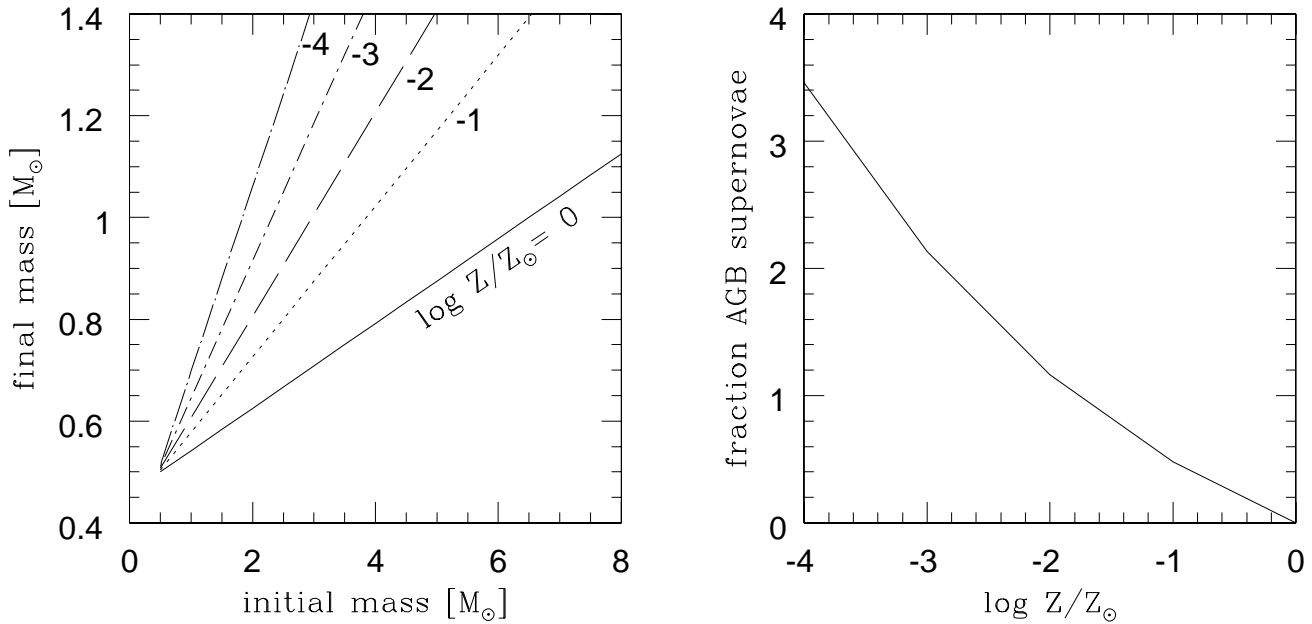}
\caption{\label{mass_relation} Left: initial-final mass relations for 
intermediate-mass
stars at low $z$. Right: The inferred ratio of AGB supernovae over
normal type-II supernovae, as function of metallicity. }
\end{figure*}

The result is shown in Fig. \ref{mass_relation}. At solar metallicity,
the most massive AGB stars reach final masses around 1\,M$_\odot$.
Already at $Z/Z_\odot=0.1$, the most massive remnants approach the
Chandresekhar limit.  [This should be taken with caution as
the most massive AGB stars may show higher luminosity than predicted
from their core masses, due to the effects of hot bottom burning
\citep{BD91}.]  At $Z/Z_\odot<0.01$, the Chandresekhar mass is reached 
for 5\,M$_\odot$ stars and at $Z/Z_\odot<0.001$, for 4\,M$_\odot$.

Carbon ignition in the degenerate core will cause these stars to
explode.  \citet{IR83} named such objects 'type 1.5 supernovae';
physically they could be considered an extension of type II SNe. The
early work did not include the superwind which was later found to
terminate the AGB before these events could occur. However, an absence
of a superwind at the lowest metallicities means that AGB supernovae
could still play a role in the early evolution of the Universe.

\section{Discussion}

\subsection{Double enriched stars}
We have presented calculations which indicate the possibility that
massive AGB stars in the early halo would have given highly massive
remnants, leading to a population of AGB supernovae. In the context of
the mixed r/s-enhanced halo stars, this suggests that the same
companion star was responsible for the double pollution.  For an
initial mass in excess of 3--4\,M$_\odot$, the companion would first
have developed into an AGB star. The wind of this star would have
carried the s-process elements. After the core reached the
Chandresekhar limit, a supernova explosion occured.  r-Process
elements formed during the explosion also polluted the companion
star. The explosion may disrupt the binary system and so this
model does not automatically predict that all such stars are
binaries. However, if a binary, the companion should be a supernova
remnant with appropriate mass.

Compared to the previously poposed models for the double enriched
stars, the present model has the advantage of not requiring a close
triple but only a binary, and it includes an explanation for the low
metallicities of the doubly enriched stars.  r-Process elements appear
to be produced in low-mass (10\,M$_\odot$) type-II supernovae
\citep{PT95}: the similar masses suggest their production could also
be expected in AGB supernovae.

\subsection{Lithium} 
\citet{SP01} find that CS 22898$-$27 has a more or less normal lithium
abundance for its metallicity, and wonder whether this is easily
explained assuming pollution from an AGB star.  AGB stars experience
lithiu-rich phases during hot bottom burning. However, the lithium is
easily destroyed again and at different phases, the surface abundances
can vary from very high to strongly depleted \citep{SB92}: the
lithium-rich phase is short-lived. The total lithium yield of low
metallicity, hot bottom burning stars can be very close to the
original Spite plateau abundance \citep{VAM02}. Thus, the normal
lithium abundance can be a coincidence. 

For hotter s-process enriched stars, the lithium 6708\AA\ line
coincides with a CeII line and derived abundances may be upper liimits
in such cases \citep{RWBQ2002}.

\subsection{Supernova rates}
The effect on the supernova rate could be significant. Assuming a
Salpeter IMF with $\alpha=-2.35$, the ratio of stars exploding as SN
on the AGB, compared to the normal type-II's is shown in
Fig. \ref{mass_relation}. At the extreme $Z$, the contribution of
these AGB-SN exceeds the normal type-II SN by a factor of a few. The
AGB stars will however explode much later and would therefore play less of
a role in the fast enrichment of the gas.

\subsection{Globular clusters}

 Old low-$Z$ population should show the effect of the higher final
masses, and contain a population of heavy white dwarfs and an
overpopulation of neutron stars.  This would affect globular clusters:
dynamical evidence for the presence of a high-$M/L$ population of
heavy remnants in the core of M15 ([Fe/H$=-2.25$) is presented by
\citet{Phinney92}, \citet{Dull97} and \citet{Baumgardt2003}. A high 
$M/L$ was also found in the core of NGC 6752 ([Fe/H$=-1.56$), which could be 
caused by an excess population of heavy white dwarfs or neutron 
stars \citep{Ferraro03}.

\section{Conclusions}

From our present knowledge of AGB mass loss, 
the early Universe may have  contained a population of supernovae which
is not found in the local, high metallicity Universe.  These AGB-SN are
possible sites for the r-process, and provide a simple explanation for
the small sample of low-metallicity, Galactic halo stars which
show evidence for both s-process and r-process enrichment.

The absence of dust driven winds at [Fe/H]$=-1$ is supported both by
models \citep{BW91} and observations (this paper). The prediction that 
stellar remnants are more massive at lower metallicities appears robust,
Regardless of whether these remnants reach the Chandresekhar limit.
The presence of high-mass white dwarfs in globular clusters is likely,
and could explain  the evidence for an excess $M/L$ in the cores
of both M 15 and NGC 6752.



\end{document}